\documentstyle[twocolumn,aps,epsfig]{revtex}

\begin{document}
\title{Entangling many atomic ensembles through laser manipulation}
\author{L.-M. Duan$^{1,2}$\thanks{
Email: lmduan@caltech.edu}}
\address{$^{1}$Institute for quantum information, mc 107-81, California
Institute of Technology, Pasadena, CA 91125-8100\\
$^{2}$Laboratory of quantum information, USTC, Hefei 230026, China}
\maketitle

\begin{abstract}
We propose an experimentally feasible scheme to generate
Greenberger-Horne-Zeilinger (GHZ) type of maximal entanglement between many
atomic ensembles based on laser manipulation and single-photon detection.
The scheme, with inherent fault tolerance to the dominant noise and
efficient scaling of the efficiency with the number of ensembles, allows to
maximally entangle many atomic ensemble within the reach of current
technology. Such a maximum entanglement of many ensembles has wide
applications in demonstration of quantum nonlocality, high-precision
spectroscopy, and quantum information processing.

\end{abstract}



\bigskip

Quantum entanglement links two or more distant subsystems in a profound
quantum mechanical way. Such a link has found wide applications in
demonstration of quantum nonlocality \cite{1,2}, high-precision spectroscopy %
\cite{3}, and quantum information processing including computation,
communication and cryptography \cite{4,5}. There are great experimental
efforts recently to get more and more subsystems entangled \cite{6,7,8,9},
since with more subsystems entangled, quantum nonlocality becomes more
striking \cite{1,2}, and the entanglement is more useful for various
applications \cite{3,4,5}. In most of the experimental \ efforts, the
subsystems are take as single-particles, and up to now three to four atoms
or photons have been entangled with a linear ion-trap \cite{8}, with a
spontaneous parametric down converter \cite{6,9}, or with a high-Q cavity %
\cite{7}. There are also proposals to entangle indistinguishable
atoms in Bose-Einstein condensates \cite{10}, or to weakly
entangle two macroscopic atomic ensembles \cite{11}, and the
latter has been demonstrated in a recent exciting experiment
\cite{12}.

In all the experimental efforts, it is hard to continuously
increase the number of the entangled subsystems due to the fast
exponential decrease of the preparation efficiency \cite{6,7,9} or
due to noise and imperfections in the setup \cite{7,8}. Here, we
propose a scheme to generate GHZ type of maximal entanglement
between many atomic ensembles with the following features:
firstly, the scheme has built-in fault-tolerance and is robust to
realistic noise and imperfections. As a result, the physical
requirements of the scheme are moderate and well fit the
experimental technique. Secondly, the preparation efficiency of
the GHZ entanglement only decreases with the number of ensembles
by a slow polynomial law. Such an efficient scaling makes it
possible to maximally entangle many (such as tens of) ensembles
with the current technology. Our scheme is based on Raman type
laser manipulation of the ensembles and single-photon detection
which postselects the desired entangled state in a probabilistic
fashion. In contrast to the belief that entangling schemes based
on postselections will necessarily suffer from the fast
exponential degradation of the efficiency, we design a scheme
which circumvents this problem by making use of quantum memory
available in atomic internal levels.

The basic element of our system is an ensemble of many identical alkali
atoms, whose experimental realization can be either a room-temperature
atomic gas \cite{12,14} or a sample of cold trapped atoms \cite{15,16}. The
relevant level structure of the atom is shown by Fig. 1. From the three
levels $\left| g\right\rangle ,\left| h\right\rangle ,\left| v\right\rangle $%
, we can define two collective atomic operators $s=\left( 1/\sqrt{N_{a}}%
\right) \sum_{i=1}^{N_{a}}\left| g\right\rangle _{i}\left\langle s\right| $
with $s=h,v,$ where $N_{a}\gg 1$ is the total \ atom number. The atoms are
initially prepared through optical pumping to the ground state $\left|
g\right\rangle $, which is effectively a vacuum state $\left| \text{vac}%
\right\rangle $ of the operators $h,v$. The $h,v$ behave like independent
bosonic mode operators as long as most of the atoms remain in the state $%
\left| g\right\rangle .$ A basis of the ``polarization'' qubit (in analogy
to the language for photons) can be defined from the states $\left|
H\right\rangle =h^{\dagger }\left| \text{vac}\right\rangle $ and $\left|
V\right\rangle =v^{\dagger }\left| \text{vac}\right\rangle $, which have an
experimentally demonstrated long coherence time \cite{12,14,15,16}.
Single-bit rotations in this basis can be done with high precision by
shining Raman pulses or radio-frequency pulses on all the atoms. The
excitations in the mode $h$ can be transferred to optical excitations \cite%
{17} and then detected by single-photon detectors. Such a transfer has a
high efficiency even for a free-space ensemble due to the collectively
enhanced coherent interaction as has been demonstrated both in \ theory \cite%
{17} and in experiments \cite{14,15}.

The first step for generation of many-party entanglement is to share an
excitation between the modes $h_{i},h_{j}$ in two distant ensembles $i,j$.
This can be readily done through a scheme in the recent quantum repeater
proposal \cite{18}, where one prepares the state $\left( h_{i}^{\dagger
}+e^{i\phi _{ij}}h_{j}^{\dagger }\right) /\sqrt{2}\left| \text{vac}%
\right\rangle $ \cite{19}, with $\phi _{ij}=\phi _{j}-\phi _{i}$, an unknown
phase difference fixed by the optical channel connecting the $i,j$
ensembles. This state, after a single-bit rotation, can be transferred to
the useful form
\begin{equation}
\left| \Psi _{i,j}\right\rangle =\left( h_{i}^{\dagger }+e^{i\phi
_{i,j}}v_{j}^{\dagger }\right) /\sqrt{2}\left| \text{vac}\right\rangle .
\end{equation}%
The basic idea of the preparation scheme in Ref. \cite{18} is as follows:
one excites the ensembles $i$ and $j$ respectively through a short weak
Raman pulse applied to the transition $\left| g\right\rangle \rightarrow
\left| e\right\rangle $\ so that the forward scattered Stokes light from the
transition $\left| e\right\rangle \rightarrow \left| h\right\rangle $ has a
mean photon number much smaller than $1$. The forward-scattered Stokes
lights from the two ensembles are then interfered at a beam splitter and
further detected by two single photon detectors. If we successfully get a
detector click, we do not know from which ensemble this registered photon
comes from, and due to this indistinguishability, the accompany collective
atomic excitation should be distributed over the ensembles $i$ and $j$ with
an equal probability amplitude, and we get the state described before. This
preparation scheme has the following two features: first, the preparation
only succeeds with a controllable small probability $p_{0}$ for each Raman
driving pulse, and needs to be repeated in average $1/p_{0}$ times for the
final successful state generation, with the total preparation time $%
t_{0}\sim 1/\left( p_{0}f_{p}\right) $, where $f_{p}$ is the repetition
frequency of the Raman pulses. Second, the scheme, with inherent resilience
to noise, is well based on the current technology of laser manipulation. We
can safely use it as our first step, to generate the state (1) with a
fidelity $F=1-p_{0}$ very close to the unity by controlling the probability $%
p_{0}$. For instance, with a typical repetition frequency $f_{p}=10$MHz, one
may prepare the state (1) with a fidelity $F=1-p_{0}\approx 99\%$ in a time $%
t_{0}\sim 10\mu $s.

Based on the preparation of the state (1), now we show how to generate
effective many-party entanglement between $n$ such atomic ensembles. We
prepare the state (1) between the $i$ and $i+1$ ensembles for each $i$ from $%
1$ to $n$, and get the following state
\begin{equation}
\left| \Psi \right\rangle =\left( 1/\sqrt{2^{n}}\right)
\prod_{i=1}^{n}\left( h_{i}^{\dagger }+e^{i\phi _{i,i+1}}v_{i+1}^{\dagger
}\right) \left| \text{vac}\right\rangle ,
\end{equation}%
where for convenience we \ have assumed the notation $n+1\equiv 1$ for the
subscripts, and have used the same symbol $\left| \text{vac}\right\rangle $
to denote the vacuum of the whole $n$ ensembles. In the expansion of the
state (1), there are only two components which have one excitation on each
ensemble. This component state is given by
\begin{equation}
\left| \Psi _{\text{eff}}\right\rangle =\left( 1/\sqrt{2}\right) \left(
\prod_{i=1}^{n}h_{i}^{\dagger }+e^{i\phi _{t}}\prod_{i=1}^{n}v_{i}^{\dagger
}\right) \left| \text{vac}\right\rangle
\end{equation}%
with $\phi _{t}=\sum_{i=1}^{n}\phi _{i,i+1}$, which is exactly the
$n$-party GHZ type maximally entangled state in the polarization
basis. Note that for any practical application of the GHZ
entanglement \cite{1,2,3,4}, the state preparation should be
succeeded by a measurement of the polarization of the excitation
on each ensemble, which can be done for our system by combining
single-bit rotations, such as Hadamard transformations, with the
number detection of the mode $h_{i}$ through single-photon
detectors. If in this measurement we only keep the results for
which an excitation appears on each ensemble (i.e., postselect the
case when the detector on each side registers a click), the states
(2) and (3) become effectively equivalent since the other
components in the state (2) have no contributions to the
measurement. Through this postselection technique, we can simply
prepare the state (2), which, whenever we put it into
applications, yields effectively the GHZ\ entanglement described
by the effective state (3). Here and in the following, we call a
component of the full state as the effective state if only this
component has contributions to the application measurements (which
are the measurements required for the detection or application of
the generated state). The effective state is the state
postselected by the application measurements.

For applications of the GHZ\ entanglement, we need also to know the phase $%
\phi _{t}$ in the effective state (3), which is fixed by the whole setup and
in principle can be measured. However, a better way is to directly cancel
this unknown phase $\phi _{t}$ with the following method. Assume that we
have an even number $n$ of the ensembles. The pair of ensembles $i$ and $%
i^{\prime }=n+2-i$ are put in near proximity so that the ensembles $i,i+1$
and $i^{\prime },\left( i+1\right) ^{\prime }$ can be connected through the
same optical channel, which fixes the phases to satisfy the relation $\phi
_{i,i+1}=\phi _{i^{\prime },\left( i+1\right) ^{\prime }}=-\phi _{\left(
i+1\right) ^{\prime },i^{\prime }}$ \cite{20}. With this relation, the
accumulated phase $\phi _{t}$ is exactly canceled to zero.

The above preparation scheme of the effective GHZ\ entanglement is
robust to realistic noise and imperfections. The dominant noise in
this system is the photon detector inefficiency, the transferring
inefficiency (induced by the spontaneous emission loss) of the
excitation from the atomic mode $h_{i}$ to the optical mode, and
the small decay of the atomic excitation in each ensemble. All the
above noise is well described by loss of excitations with a
overall loss probability denoted by $\eta $. Note that by
including the detector inefficiency, we have automatically taken
into account that the single-photon detectors cannot perfectly
distinguish between single and two photons. It is easy to see that
loss of excitations only has influence on the success probability
to register an excitation from each ensemble. Whenever the
excitation is registered, its polarization is still perfectly
entangled as shown by the effective state (3).

Now we consider the efficiency of this scheme, which can be described by the
total time needed to successfully register the effective GHZ entanglement.
The preparation of the factor state (1) is probabilistic, however, due to
the available quantum memory provided by the metastable atomic modes $h,v$,
the preparation time $t_{1}$ of the state (2)\ is at most $nt_{0}$ if its
factor states are prepared one after the other, and can be reduced to $%
t_{1}\sim t_{0}$ (in the case of $n<1/p_{0}$) if its factor states are
prepared independently at the same time. In contrast to this, in the case of
no quantum memory, one would need about $1/p_{0}^{n}$ repeats of the Raman
pulses for a successful preparation of the state (2), and a total time $%
t_{0}/p_{0}^{n-1}\gg t_{0}$. After preparation of the state (2), the
projection efficiency (success probability) from the state (2) to the
effective GHZ state (3) is given by $\left( 1-\eta \right) ^{n}/2^{n-1}$,
where we have assumed the same loss probability $\eta $ for each ensemble.
So the total time for registering the $n$-party GHZ entanglement is $T\sim
t_{0}2^{n-1}/\left( 1-\eta \right) ^{n}$, which increases with the number of
ensembles exponentially by the factor $2/\left( 1-\eta \right) $. Note that
this increase has been much slower than the case for spontaneous parametric
down conversion where the exponential increasing factor is about $2$ orders
larger due to the absence of quantum memory \cite{6,9}.

We can in fact further improve the scheme to get a much more efficient
scaling of the efficiency, with the time $T$ increasing with the party
number $n$ only polynomially. The improved scheme is divided into the
following three steps:

(i) We start with two pairs of ensembles $1,2$ and $3,4$, prepared in the
state $\left| \Psi _{1,2}\right\rangle \bigotimes \left| \Psi
_{3,4}\right\rangle $ with $\left| \Psi _{i,j}\right\rangle $ in the form of
Eq. (1). We then connect these two disjoint pairs by preparing the state $%
\left| \Psi _{2,3}\right\rangle $. The ensembles $2$ and $3$ will
not be involved any more in the following steps for state
preparation, so we can immediately put them into applications by
doing the same type of measurements on them as if we had generated
$n$-party GHZ\ entanglement. In these measurements, if one
excitation is registered from each ensemble $2$ and $3$, we
succeed and will go on with the next step. Otherwise, we simply
repeat the above process until we succeed. Upon success, only the component $%
\left| \Psi _{1-4}\right\rangle $ of the state $\left| \Psi
_{1,2}\right\rangle \bigotimes \left| \Psi _{3,4}\right\rangle \bigotimes
\left| \Psi _{2,3}\right\rangle $ has contributions to the measurement with
\begin{equation}
\left| \Psi _{1\text{-}4}\right\rangle =\left( 1/\sqrt{2}\right) \left(
h_{1}^{\dagger }h_{2}^{\dagger }h_{3}^{\dagger }+v_{2}^{\dagger
}v_{3}^{\dagger }v_{4}^{\dagger }\right) \left| \text{vac}\right\rangle ,
\end{equation}%
where for simplicity we have neglected the phase $\phi _{i,i+1}$ since they
will finally cancel each other with the method described before. If loss of
excitations with a loss probability $\eta $ is taken into account for
detections on the ensembles $2,3$, a registered click might result from two
excitations, and in this case there will be no excitation in the ensembles $1
$ and $4$. So with the loss, upon success of step (i), instead of $\left|
\Psi _{1\text{-}4}\right\rangle $ the effective state of the ensembles 1-4
is actually described by
\begin{equation}
\rho _{1\text{-}4}=\left( \left| \Psi _{1\text{-}4}\right\rangle
\left\langle \Psi _{1\text{-}4}\right| +c_{1}\rho _{\text{vac}}\right)
/\left( 1+c_{1}\right)
\end{equation}%
with the vacuum coefficient $c_{1}=2\eta $, where $\rho _{\text{vac}}$
stands for the vacuum component with no excitation in the undetected
ensembles $1$ and $4$. The probability of a successful detection on both of
the ensembles $2$ and $3$ is given by $p_{1}=\left( 1-\eta \right)
^{2}\left( 1+2\eta \right) /4$, which means that we need to repeat the
process in average $1/p_{1}$ times for the final success of step (i).

(ii) In step (ii) we further extend the number of entangled ensembles in the
effective state (5). Assume that we have applied the method of step (i) in
parallel to the two disjoint sets of ensembles $1$-$4$ and $5$-$8$, with
their effective states (each in the form of Eq. (5)) denoted by $\rho _{1%
\text{-}4}$ and $\rho _{5\text{-}8}$, respectively. We connect these two
sets by first preparing the state $\left| \Psi _{4,5}\right\rangle $ (in the
form of Eq. (1)) and then putting the ensembles $4,5$ into application
measurements as described in step (i). Upon success of these measurements
with one excitation registered from each ensemble, the postselected state of
the ensembles $1$-$8$ is effectively described by $\rho _{1\text{-}8}$ which
is similar to Eq. (5), but with an increased vacuum coefficient and with $%
\left| \Psi _{1\text{-}4}\right\rangle $ replaced by $\left| \Psi _{1\text{-}%
k}\right\rangle =\left( 1/\sqrt{2}\right) \left(
\prod_{i=1}^{k-1}h_{i}^{\dagger }+\prod_{i=2}^{k}v_{i}^{\dagger }\right)
\left| \text{vac}\right\rangle ,$ $\left( k=8\right) $. Whenever the
measurement fails, we repeat the whole state preparation from step (i). The
above connection process can be continued with the number $n$ of effectively
entangled ensembles doubled for each time of connection. After $i$ times
connection, we have $n=2^{i+1}$. The success probability and the new vacuum
coefficient of the $i$th connection are denoted respectively by $p_{i}$ and $%
c_{i}$, which satisfy the following recursion relations with the previous
vacuum coefficient $c_{i-1}$ through $p_{i}=\left( 1-\eta \right) ^{2}\left(
1+2\eta +2c_{i-1}\right) /\left[ 4\left( 1+c_{i-1}\right) ^{2}\right] $, and
$c_{i}=2c_{i-1}+2\eta $. From these recursion relations, we have $%
c_{i}=2\eta \left( 2^{i}-1\right) $, which, after substituted into $p_{i}$,
yields an explicit expression for the repetition number $1/p_{i}$ of the $i$%
th connection.

(iii) After a desired number $n=2^{i+1}$ of the ensembles have been
entangled in the effective state $\rho _{1\text{-}n}$, we close the loop in
the last step by first preparing the state $\left| \Psi _{n,1}\right\rangle $
(in the form of Eq. (1)) and then putting the last two ensembles $n,1$ into
application measurements. As usual, we keep the results only when one
excitation appears from each detected ensemble, and this automatically
eliminates contributions from the vacuum component in the state $\rho _{1%
\text{-}n}$. So the effective state of the whole set of ensembles
postselected by all the application measurements is still described by the
exact GHZ state (3), and the application measurement results should reveal
perfect GHZ\ entanglement between the $n$ ensembles in the polarization
degree of freedom. The application measurements on the ensembles $n,1$ in
the last step succeeds with a probability $p_{l}=\left( 1-\eta \right) ^{2}/%
\left[ 2\left( 1+c_{i}\right) \right] $, so the whole process needs to be
repeated in average $1/p_{l}$ times.

Now we calculate in this improved scheme how much time is needed in total
for a successful detection of the $n$-party GHZ\ entanglement. This time is
given by $T_{\text{imp}}=t_{0}/\left[ p_{l}p_{1}\prod_{j=2}^{i}p_{j}\right] $%
, with $t_{0}$, the preparation time of the state (1). We consider two
limiting cases. In the first case with a negligible loss probability $\eta $
for each ensemble, we have $p_{l}=2p_{j}=1/2$ and $T_{\text{imp}%
}=2^{2i+1}t_{0}=n^{2}t_{0}/2$, which increases with the number $n$ of
entangled ensembles by the slow quadratic law. In the second case with a
considerably large loss probability $\eta $, the total time $T_{\text{imp}}$
is approximated by $T_{\text{imp}}\sim t_{0}\left[ 2\eta n/\left( 1-\eta
\right) ^{2}\right] \left( n/2\right) ^{\log _{2}\left[ 2\eta \sqrt{n}%
/\left( 1-\eta \right) ^{2}\right] }$, which increases with $n$ faster, but
still polynomially (or, more accurately, sub-exponentially). The basic
reason for the improvement from the exponential scaling to the much slower
polynomial scaling is due to that we have divided the whole preparation
process into many small steps, checking in each step whether the preparation
is successful, and repeating this small step instead of the whole process if
it fails.

Finally, we briefly discuss the practical implication of this proposal. With
the improved scheme, for example, we can generate high-fidelity GHZ
entanglement over $n=16$ ensembles in a time $T_{\text{imp}}\sim 50$ms with
a notable loss $\eta \approx 1/3$ and a typical choice $t_{0}\sim 10\mu $s.
With such a short preparation time $T_{\text{imp}}$, the noise that we have
not included, such as the non-stationary phase drift induced by the pumping
lase or by the optical channel, is negligible. As long as the number $n$ of
the ensembles is not huge, we can also safely neglect the single-bit
rotation error (below $10^{-4}$ with the use of accurate polarization
techniques for Zeeman sublevels \cite{21} and the dark count probability
(about $10^{-5}$ in a typical detection time window $0.1\mu $s) of
single-photon detectors. Due to the efficient scaling of this scheme, one
can use it to steadily increase the number of entangled ensembles, and it
seems reasonable to generate GHZ entanglement over tens of ensembles with
the current technology. Such an extraordinary possibility opens up prospects
for many exciting experiments and applications.

{\bf Acknowledgments} This work was supported in part by the Caltech MURI
Center for Quantum Networks under ARO Grant No. DAAD19-00-1-0374, by the
National Science Foundation under Grant No. EIA-0086038, and also by the
Chinese Science Foundation and Chinese Academy of Sciences.

\begin{figure}[tbp]
\epsfig{file=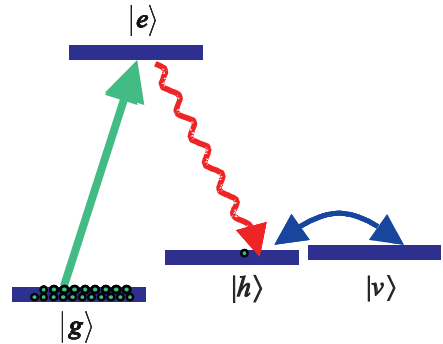,width=12cm}
 \caption{The relevant
atomic level structure with $\left| g\right\rangle $, the ground
state, $\left| e\right\rangle $ , the excited state, and $\left|
h\right\rangle ,$ $\left| v\right\rangle $ the two metastable
states (e.g., Zeeman or hyperfine sublevels) for storing a qubit
of information. The three levels $\left| g\right\rangle $ ,
$\left| e\right\rangle $ , and $\left| h\right\rangle $ can be
coupled through a Raman process which is useful for measurement of
the collective atomic excitation in the state $\left|
h\right\rangle $ [16] and for generating preliminary entanglement
between two ensembles [17].}
\end{figure}


\begin{references}
\bibitem{19} Ref. [17] shows that the generated state will be mixed with a
small vacuum component if we take into account the detector dark
counts. However, this vacuum component is typically very small
since the normal dark count rate (100 Hz) is much smaller than the
repetition frequency (10 MHz, for instance) of the Raman pulses,
and in addition, this component will be finally automatically
eliminated in our scheme since its effect is the same as the
effect of detector inefficiency in the application measurements
which we will analyze below. So, for simplicity, we write a pure
state here by neglecting this small component.
\bibitem{20} Remember the notation $n+1\equiv 1$. The index $i$ and $%
i^{\prime }=n+2-i$ actually denote the same ensemble for $i=1$ and $n/2+1.$
So as a special case of the relation $\phi _{i,i+1}=-\phi _{\left(
i+1\right) ^{\prime },i^{\prime }}$, we have $\phi _{1,2}=-\phi _{n,1}$ and $%
\phi _{n/2,n/2+1}=-\phi _{n/2+1,n/2+2}$.
\end{references}
\end{document}